\documentclass[twocolumn,aps,prl,showpacs]{revtex4}
\usepackage{graphicx}
\usepackage{amsmath,amssymb}
\def\endproof{\vrule height6pt width6pt depth0pt}

\begin{document}
\title{Bipartite All-Versus-Nothing Proofs of Bell's Theorem with Single-Qubit Measurements}
\author{Ad\'{a}n Cabello}
\email{adan@us.es}
\affiliation{Departamento de F\'{\i}sica Aplicada II,
Universidad de Sevilla, E-41012 Sevilla, Spain}
\author{Pilar Moreno}
\affiliation{Departamento de F\'{\i}sica Aplicada II,
Universidad de Sevilla, E-41012 Sevilla, Spain}
\date{\today}

%%%%%%%%%%%%%%%%%%%%%%%%%%%%%%%%%%%%%%%%%%%%%%%%%%%%%%%%%%%%%%%%%%%

%First version: 11 January 2007
%This version: 18 November 2007 (after PRL's proofs)

%%%%%%%%%%%%%%%%%%%%%%%%%%%%%%%%%%%%%%%%%%%%%%%%%%%%%%%%%%%%%%%%%%%

\begin{abstract}
If $n$ qubits were distributed between 2 parties, which quantum pure
states and distributions of qubits would allow all-versus-nothing
(or Greenberger-Horne-Zeilinger-like) proofs of Bell's theorem using
only single-qubit measurements? We show a necessary and sufficient
condition for the existence of these proofs for {\em any} number of
qubits, and provide {\em all} distinct proofs up to $n=7$~qubits.
Remarkably, there is only one distribution of a state of $n=4$
qubits, and six distributions, each for a different state of $n=6$
qubits, which allow these proofs.
\end{abstract}

%%%%%%%%%%%%%%%%%%%%%%%%%%%%%%%%%%%%%%%%%%%%%%%%%%%%%%%%%%%%%%%%%%%

%Optical tests of quantum theory
\maketitle

%%%%%%%%%%%%%%%%%%%%%%%%%%%%%%%%%%%%%%%%%%%%%%%%%%%%%%%%%%%%%%%%%%%

The Greenberger-Horne-Zeilinger (GHZ) proof~\cite{GHZ89,Mermin90a}
of Bell's theorem~\cite{Bell64} not only ``opened a new chapter on
the hidden variables problem'' \cite{Zukowski91} and made ``the
strongest case against local realism since Bell's work''
\cite{CRB91}; it also inspired the quantum protocols for reducing
communication complexity \cite{CB97} and for secret sharing
\cite{ZZHW98}, and motivated the study of multipartite entanglement
\cite{DVC00}. The GHZ proof provides a direct contradiction, using
qubits and without requiring inequalities, between Einstein,
Podolsky, and Rosen's (EPR's) criterion of elements of reality
\cite{EPR35} and perfect correlations predicted by quantum
mechanics. Mermin coined the name ``all-versus-nothing'' (AVN) for
proofs like GHZ's, based on $m$ perfect correlations such that, if
we assume elements of reality, $m-1$ of them lead us to the
conclusion that it is the {\em opposite} of the one given by the
$m$th correlation \cite{Mermin90b}.

However, while the original proof of Bell's theorem required only 2
separated parties, the GHZ proof required 3 because, when the qubits
are distributed between 2 parties, there is no physical reason
supporting the assumption that all single-qubit observables
appearing in the proof have predefined results, since some of them
do {\em not} satisfy EPR's criterion of elements of reality. EPR's
criterion states that: ``if, without in any way disturbing a system,
we can predict with certainty (i.e., with probability equal to
unity) the value of a physical quantity, then there exists an
element of physical reality corresponding to this physical
quantity'' \cite{EPR35}. Applied to the bipartite case, this means
that it must be possible to predict with certainty the results of
measuring all observables appearing in the proof on Alice's (Bob's)
side from the results of spacelike separated measurements on Bob's
(Alice's) side.

The first 2-party AVN proof with qubits was introduced in
\cite{Cabello01a,Cabello01b}, then adapted for 2 photons
\cite{CPZBZ03}, and finally tested in the
laboratory~\cite{CBPMD05,YZZYZZCP05}. One of the difficulties of
experimentally implementing this 2-party AVN proof was that it
required 2-qubit local measurements \cite{Lvovsky02}. The first
2-party AVN proof requiring only single-qubit measurements was
introduced in \cite{Cabello05a,Cabello05b} and has been recently
demonstrated in the laboratory \cite{VPMDB07}. These bipartite AVN
proofs required 4-qubit states with 2 qubits each on Alice's and
Bob's sides.

The possibilities brought forth by recent developments like 2-photon
hyperentangled states (i.e., entangled in several degrees of
freedom) encoding 3 or more qubits in each photon \cite{BLPK05}, and
6-photon 6-qubit states \cite{ZGWCZYMSP06,LZGGZYGYP07}, naturally
lead to the following problem: If $n$ qubits were distributed
between 2 parties, which are the quantum pure states and possible
distributions of qubits that allow a 2-party AVN proof using only
single-qubit measurements?

This problem is also related to the one of finding genuinely new
bipartite communication complexity problems with a quantum advantage
(specifically, new schemes of quantum pseudotelepathy
\cite{Cabello06a}), and to the problem of deciding which $n$-qubit
states and distributions of qubits allow bipartite EPR-Bell
inequalities \cite{Cabello06b,BDMVC06}.

In this Letter we show a necessary and sufficient condition for the
existence of bipartite AVN proofs using only single-qubit
measurements (BAVN hereafter) for {\em any} number of qubits. We
then proceed to explicitly provide {\em all} physically distinct
BAVN proofs with up to 7~qubits.

A BAVN proof consists of an $n$-qubit quantum state and a set of
single-qubit measurements that satisfy two requirements:
(a) {\em Perfect correlations to define bipartite EPR's elements of
reality.} Every single-qubit observable involved in the proof must
satisfy EPR's criterion of elements of reality.
(b) {\em Perfect correlations that contradict EPR's elements of
reality.} The observables that satisfy EPR's condition {\em cannot}
have predefined results, because it must be impossible to assign
them values that satisfy all the perfect correlations predicted by
quantum mechanics.

Perfect correlations are necessary to establish elements of reality
and to prove that they are incompatible with quantum mechanics.
Therefore, the states we are interested in must be simultaneous
eigenstates of a sufficient number of commuting $n$-fold tensor
products of single-qubit operators. Suppose that $A$ and $B$ are
single-qubit operators on the same qubit. If they are different,
they cannot be commuting operators. The only way to make the
$n$-fold tensor products be commuting operators is to choose $A$ and
$B$ to be {\em anticommuting} operators. Therefore, in an AVN proof,
all the local operators corresponding to the same qubit must be
anticommuting operators. The maximum number of anticommuting
single-qubit operators is 3. Therefore, without loss of generality,
we can restrict our attention to a specific set of 3 single-qubit
anticommuting operators on each qubit, e.g. the Pauli matrices
$X=\sigma_x$, $Y=\sigma_y$, and $Z=\sigma_z$. This leads us to the
notion of stabilizer states. An $n$-qubit {\em stabilizer state} is
defined as the simultaneous eigenstate with eigenvalue $1$ of a set
of $n$ independent (in the sense that none of them can be written as
a product of the others) commuting elements of the Pauli group,
defined as the group, under matrix multiplication, of all $n$-fold
tensor products of $X$, $Y$, $Z$, and the identity $\openone$. The
$n$ independent elements are called {\em stabilizer generators} and
generate a maximally Abelian subgroup called the {\em stabilizer
group} of the state \cite{Gottesman96}. The $2^n$ elements of the
stabilizer group are called {\em stabilizing operators}, and provide
all the prefect correlations of the stabilizer state.

Moreover, since any stabilizer state is local Clifford equivalent
(i.e., equivalent under the local unitary operations that map the
Pauli group to itself under conjugation) to a graph state
\cite{VDM04}, then we can restrict our attention to graph states. A
{\em graph state} \cite{HEB04} is a stabilizer state whose
generators can be written with the help of a graph. $|G\rangle$ is
the $n$-qubit state associated to the graph $G$, which gives a
recipe both for preparing $|G\rangle$ and for obtaining $n$ stabilizer
generators that uniquely determine $|G\rangle$. On one hand, $G$ is
a set of $n$ vertices (each of them representing a qubit) connected
by edges (each of them representing an Ising interaction between the
connected qubits). On the other hand, the stabilizer generator $g_i$
is obtained by looking at the vertex $i$ of $G$ and the set $N(i)$
of vertices which are connected to $i$, and is defined by
\begin{equation}
g_i = X_i \bigotimes_{j \in N(i)} Z_j, \label{grule}
\end{equation}
where $X_i$, $Y_i$, and $Z_i$ denote the Pauli matrices acting on
the $i$th qubit. $|G\rangle$ is the unique $n$-qubit state that
fulfills
\begin{equation}
g_i |G\rangle = |G\rangle, \text{ for }i=1,\ldots,n. \label{erule}
\end{equation}
Therefore, the stabilizer group is
\begin{equation}
S(|G\rangle)=\{s_j, j=1,\ldots,2^n\}; \;\;\; s_j=\prod_{i \in
I_j(G)} g_i, \label{stabilizer}
\end{equation}
where $I_j(G)$ denotes a subset of $\left\{g_i\right\}_{i=1}^N$. The
stabilizing operators of $|G\rangle$ satisfy
\begin{equation}
s_j |G\rangle = |G\rangle. \label{stabilizingoperatorseqs}
\end{equation}
Equations like (\ref{stabilizingoperatorseqs}) are the ones that can
be used to establish elements of reality and prove their
incompatibility with quantum mechanics.

Although graph states are now ubiquitous in quantum information
theory due to their role as code words of quantum error correcting
codes \cite{Gottesman96}, or in measurement-based quantum
computation \cite{RB01}, or due to their use in the classification
of entanglement \cite{HDERVB06}, the first $n>2$-qubit graph states
were the GHZ states and appeared in the context of AVN proofs. It is
then not that surprising that, when we want to obtain BAVN proofs,
we go back to graph states. Indeed, DiVincenzo and Peres already
showed that the requirement (b) does not only occur for GHZ states,
but is also inherent to all standard code words of quantum error
correcting codes \cite{DP97}. More recently, Scarani {\em et al.}
have shown that (b) holds for cluster states constructed on square
lattices of any dimension \cite{SASA05}. Furthermore, a positive
by-product of focusing on graph states is that graph states
associated to connected graphs have been exhaustively classified.
There is only one 2-qubit graph state (equivalent to a Bell state),
only one 3-qubit graph state (the GHZ state), two 4-qubit graph states
(the GHZ and the cluster state), four 5-qubit graph states, eleven 6-qubit
graph states, and twenty-six 7-qubit graph states \cite{HEB04}.

Therefore, our problem reduces to the following: If $n$ qubits were
distributed between 2 parties, which $n$-qubit graph states and
possible distributions of qubits allow a bipartite AVN proof using
only single-qubit observables?

Note that, even considering only up to $7$ qubits, there are
hundreds of states and possible distributions that could
potentially lead to a BAVN proof. Remarkably, this is not the case.

Our starting point is the observation that requirement (b) is
satisfied by {\em any} graph state.

{\em Lemma~1.}---Any graph state associated to a connected graph of
$3$ or more vertices leads to algebraic contradictions with the
concept of elements of reality (when each qubit is distributed to a
different party).

This result was anticipated in \cite{DP97,SASA05,HDERVB06}. The
interest of the following proof is that it provides methods for
obtaining explicit examples of sets of perfect correlations
satisfying (b).

{\em Proof}.---If qubit~$i$ is connected to qubit~$j$, and $j$ is
connected to $k$, there are two possibilities. One is that $i$ is
not connected to $k$. Then, no theory exists that assigns
predefined values $-1$ or $1$ to $Y_i$, $Z_i$, $X_j$, $Y_j$, $Y_k$,
and $Z_k$ simultaneously satisfying the four equations
\begin{subequations}
\begin{align}
g_i g_j |G\rangle = |G\rangle, \\
g_j |G\rangle = |G\rangle, \\
g_j g_k |G\rangle = |G\rangle, \\
g_i g_j g_k |G\rangle = |G\rangle,
\end{align}
\end{subequations}
since $g_i g_j \cdot g_j \cdot g_j g_k$ (where ``$\cdot$'' means
matrix multiplication) is equal, not to $g_i g_j g_k$ (as expected
in any theory with predefined values), but to $-g_i g_j g_k$.

The other possibility is that qubit~$i$ is also connected to $k$.
Then, no theory exists that assigns predefined values $-1$ or $1$
to $X_i$, $Z_i$, $X_j$, $Z_j$, $X_k$, and $Z_k$ simultaneously
satisfying the four equations
\begin{subequations}
\begin{align}
g_i |G\rangle = |G\rangle, \\
g_j |G\rangle = |G\rangle, \\
g_k |G\rangle = |G\rangle, \\
g_i g_j g_k |G\rangle = |G\rangle,
\end{align}
\end{subequations}
since $g_i \cdot g_j \cdot g_k$ is equal to $-g_i g_j
g_k$.\hfill\endproof

Any set of equations associated to the stabilizing operators
containing a subset satisfying (b) also satisfies (b). Therefore,
given a graph state associated to a connected graph of $n
>3$ vertices, there are {\em thousands} of possible different subsets
of equations satisfying (b). Most of them involve the three Pauli
matrices of all the qubits, but some of them do not. However, in our
BAVN proofs it is relevant that the three Pauli matrices of {\em
each and every one} of Alice's (Bob's) qubits can be regarded as EPR
elements of reality, because we are interested in new BAVN proofs
involving new classes of graph states, not those which are mere
consequences of previously considered graph states of fewer qubits.

Therefore, the problem we have to solve is that of finding out for
which graph states and distributions are all the three Pauli
matrices for all the single-qubit elements of reality in a
bipartite scenario. A distribution of $n$ qubits between Alice and
Bob is said to permit {\em bipartite elements of reality} when, for
each and every qubit, the results of measuring two Pauli matrices on
Alice's (Bob's) qubit~$j$ can be predicted with certainty from the
results of measurements on Bob's (Alices's) qubits only.

Let us define the {\em reduced stabilizer} of Alice's (Bob's) qubits
as the one obtained by tracing out Bob's (Alice's) qubits. A
necessary and sufficient condition for bipartite elements of reality
is the following.

%%%%%%%%%%%%%%%%%%%%%%%%%%%%%%%%%%%%%%%%%%%%%%%%%%%%%%%%%%%%%%%%%%%

{\em Lemma~2.}---A distribution of $n$ qubits between Alice (who is
given $n_A$ qubits) and Bob (who is given $n_B=n-n_A$ qubits)
permits bipartite elements of reality if and only if $n_A=n_B$, and
the reduced stabilizer of Alice's (Bob's) qubits contains {\em all}
possible variations with repetition of the four elements,
$\openone$, $X$, $Y$, and $Z$, choose $n_A$ ($n_B$), without
repeating any of them.

%%%%%%%%%%%%%%%%%%%%%%%%%%%%%%%%%%%%%%%%%%%%%%%%%%%%%%%%%%%%%%%%%%%

{\em Proof.}---Suppose that two Pauli matrices of Alice's qubit~$1$,
e.g. $X_1$ and $Y_1$ are elements of reality. Then each of them must
be predicted with certainty from Bob's measurements. That is, the
reduced stabilizer of Alice's qubits must contain
\begin{subequations}
\begin{align}
X_1 \otimes \openone_2 \otimes \ldots \otimes \openone_{n_A}, \label{x1} \\
Y_1 \otimes \openone_2 \otimes \ldots \otimes \openone_{n_A}.
\label{y1}
\end{align}
\end{subequations}
Therefore, the third Pauli matrix of Alice's qubit~$1$ must also be
an element of reality, since the product of (\ref{x1}) and
(\ref{y1}), which must belong to the reduced stabilizer of Alice's
qubits, is
\begin{equation}
Z_1 \otimes \openone_2 \otimes \ldots \otimes \openone_{n_A}.
\label{z1}
\end{equation}
The same must happen with the three Pauli matrices of Alice's qubits
$2, \ldots, n_A$. Therefore, the reduced stabilizer of Alice's
qubits must also contain
\begin{subequations}
\begin{align}
& \openone_1 \otimes X_2 \otimes \openone_3 \otimes \ldots \otimes
\openone_{n_A}, \label{x2} \\
& \openone_1 \otimes Y_2 \otimes \openone_3 \otimes \ldots \otimes
\openone_{n_A},
\label{y2} \\
& \openone_1 \otimes Z_2 \otimes \openone_3 \otimes \ldots \otimes
\openone_{n_A}, \ldots,
\label{z2} \\
& \openone_1 \otimes \ldots \otimes \openone_{n_A-1} \otimes Z_{n_A}.
\label{znA}
\end{align}
\end{subequations}
Moreover, the reduced stabilizer of Alice's qubits must contain all
the possible products of the Eqs. (\ref{x1})--(\ref{znA}); that is,
all possible variations with repetition of the four elements,
$\openone$, $X$, $Y$, and $Z$, choose $n_A$, which are $4^{n_A}=2^{2
n_A}$. Furthermore, a similar reasoning applies to the three Pauli
matrices of each and every one of Bob's qubits. Therefore, the
reduced stabilizer of Bob's qubits must also contain all the
possible products of
\begin{subequations}
\begin{align}
& X_{n_A+1} \otimes \openone_{n_A+2} \otimes \ldots \otimes \openone_{n_B}, \ldots, \label{x1B} \\
& \openone_{n_A+1} \otimes \ldots \otimes \openone_{n_B-1} \otimes
Z_{n_B}. \label{znB}
\end{align}
\end{subequations}
But the total stabilizer only has $2^{n_A+n_B}$ terms; therefore the
only possibility is that $n_A = n_B$. In addition, note that there
is no space for any of the variations with repetition to be
repeated.\hfill\endproof

Most of the graph states {\em cannot} be used in BAVN proofs. The
remarkable point is that there are a few graph states and
distributions of qubits that satisfy the requirements of Lemma~2,
and therefore simultaneously fulfill (a) and (b). Moreover, since
Lemma~2 is a necessary and sufficient condition, when we apply it to
every possible distribution of qubits of all possible graph states,
we obtain a {\em complete} classification of all possible BAVN
proofs.

%%%%%%%%%%%%%%%%%%%%%%%%%%% Figure 1 %%%%%%%%%%%%%%%%%%%%%%%%%%%%

\begin{figure*}
\centerline{\includegraphics[width=1.28 \columnwidth]{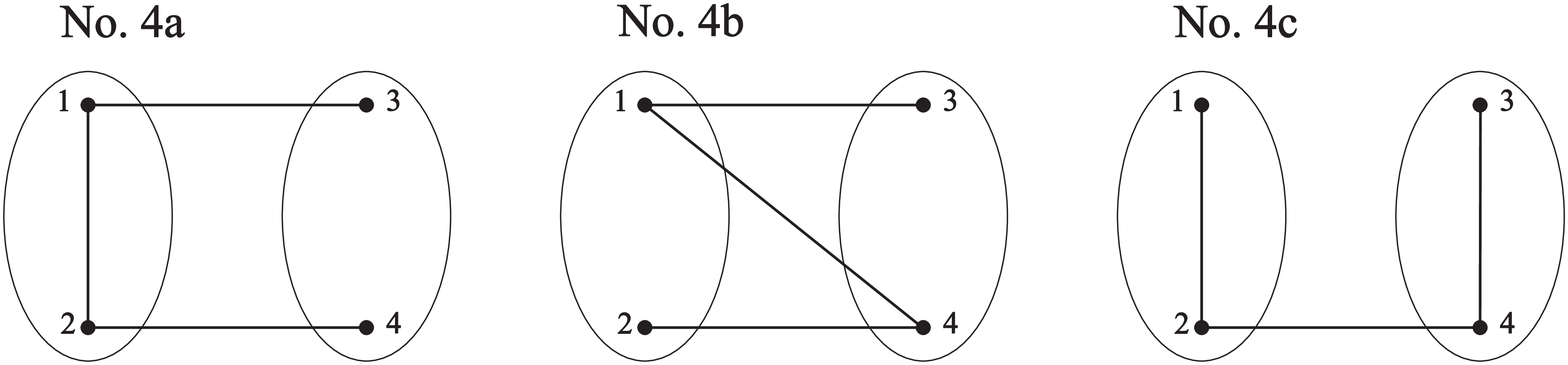}}
\caption{\label{Fig1}Bipartite distributions of the 4-qubit cluster
state (graph state no.~4 according to Hein {\em et al.}
\cite{HEB04}). Distribution 4a permits bipartite elements of reality
and BAVN proofs. Distribution 4b is physically equivalent (it is
just relabeling the basis). Distribution 4c is not equivalent to the
other two, and does not permit bipartite elements of reality.}
\end{figure*}

%%%%%%%%%%%%%%%%%%%%%%%%%%%%%%%%%%%%%%%%%%%%%%%%%%%%%%%%%%%%%%%%%

With $n < 8$ qubits, and modulo single-qubit unitary
transformations, the only states and distributions of qubits that
allow BAVN proofs are the following. There is only one graph state
with $4$~qubits:
\begin{equation}
|\psi_{4a}\rangle=\frac{1}{2}(|00\rangle|\bar0\bar0\rangle+
|01\rangle|\bar0\bar1\rangle+
|10\rangle|\bar1\bar0\rangle-
|11\rangle|\bar1\bar1\rangle),
\end{equation}
where $|00\rangle|\bar0\bar0\rangle = |\sigma_z=0\rangle_1 \otimes
|\sigma_z=0\rangle_2 \otimes |\sigma_x=0\rangle_3 \otimes
|\sigma_x=0\rangle_4$, with qubits $1$ and $2$ in Alice's side, and
qubits $3$ and $4$ in Bob's. The state $|\psi_{4a}\rangle$
corresponds to the graph state no.~4 according to Hein {\em et al.}
\cite{HEB04}, with its qubits distributed as in Fig.~\ref{Fig1},
distribution 4a. Note that any other non-equivalent distribution of
qubits does not allow BAVN proofs (see Fig.~\ref{Fig1}). This BAVN
proof is precisely the one introduced in \cite{Cabello05a}. The new
result is that the proof in \cite{Cabello05a} is {\em the only one}
with 4 qubits and single qubit measurements.

%%%%%%%%%%%%%%%%%%%%%%%%%%% Figure 2 %%%%%%%%%%%%%%%%%%%%%%%%%%%%

\begin{figure}
\centerline{\includegraphics[width=0.84\columnwidth]{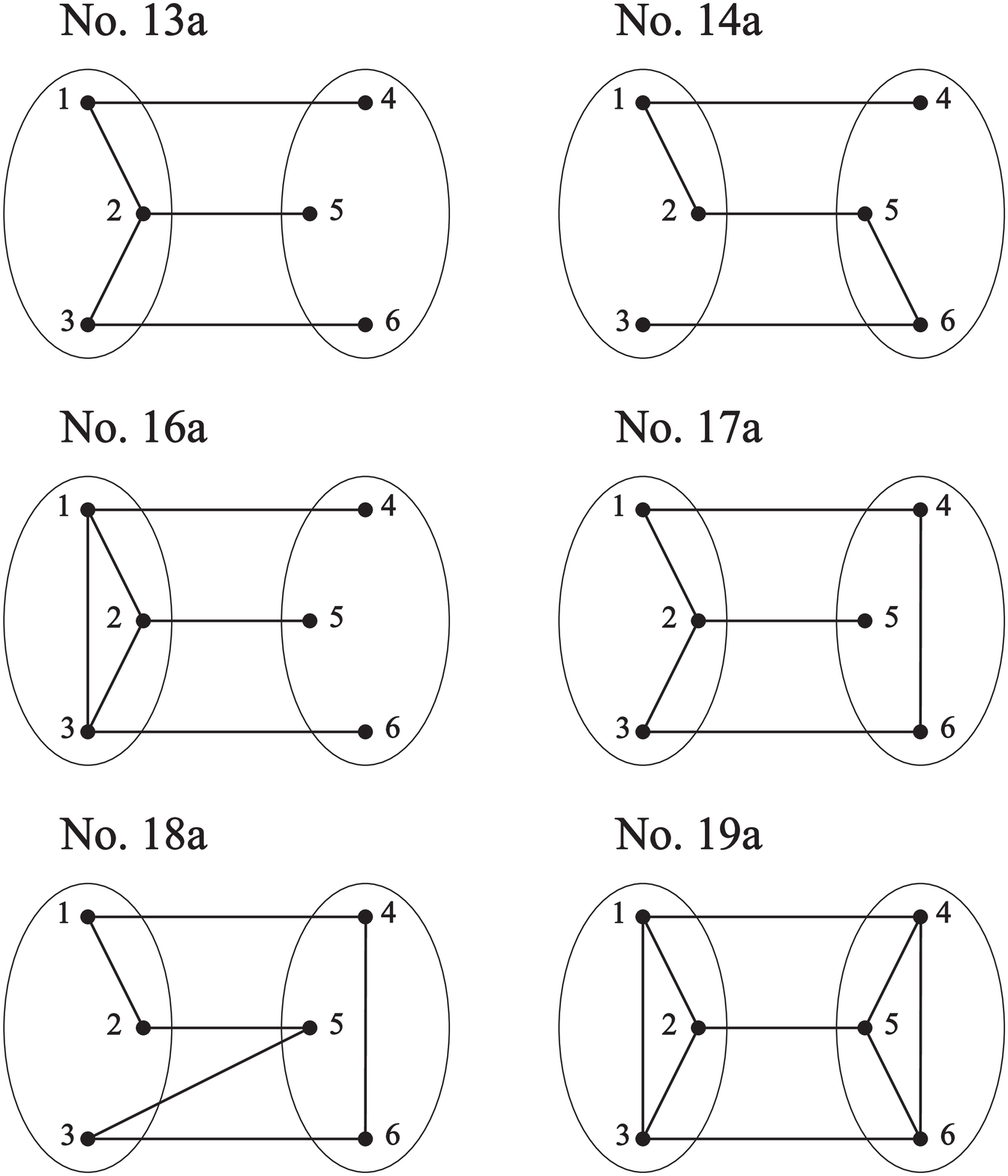}}
\caption{\label{Fig2}Bipartite distributions of the 6-qubit graph
states that permit bipartite elements of reality and BAVN proofs.
The graphs' nomenclature follows Hein {\em et al.} \cite{HEB04}, but
the labeling of the qubits is different: Qubits $1$, $2$, and $3$
belong to Alice, and qubits $4$, $5$, and $6$ belong to Bob.}
\end{figure}

%%%%%%%%%%%%%%%%%%%%%%%%%%%%%%%%%%%%%%%%%%%%%%%%%%%%%%%%%%%%%%%%%

Between $5$ and $7$ qubits, there are only $6$ possible states and
distributions leading to BAVN proofs. All of them are $6$-qubit
states in which each party has $3$ qubits. Their corresponding
graphs are summarized in Fig.~\ref{Fig2}. The explicit expressions
of each state can be obtained from its graph using (\ref{grule}) and
(\ref{erule}). Two $6$-qubit graph states have been recently
prepared in the laboratory \cite{ZGWCZYMSP06,LZGGZYGYP07}, but none
of them allows BAVN proofs. A $6$-qubit BAVN proof constitutes an
interesting experimental challenge for the near future.

%%%%%%%%%%%%%%%%%%%%%%%%%%%%%%%%%%%%%%%%%%%%%%%%%%%%%%%%%%%%%%%%%%%

%\section*{Acknowledgments}

%%%%%%%%%%%%%%%%%%%%%%%%%%%%%%%%%%%%%%%%%%%%%%%%%%%%%%%%%%%%%%%%%%%

This work was sparked by questions made by F. Verstraete and M.
\.{Z}ukowski. The authors thank H. Briegel, W. D\"{u}r, O.
G\"{u}hne, A. J. L\'{o}pez-Tarrida, and M. van den Nest for their
useful comments, and acknowledge support by the Spanish MEC Project
No. FIS2005-07689, and the Junta de Andaluc\'{\i}a Excellence
Project No. P06-FQM-2243.

%%%%%%%%%%%%%%%%%%%%%%%%%%% References %%%%%%%%%%%%%%%%%%%%%%%%%%%%

%%%%%%%%%%%%%%%%%%%%%%%%%%%%%%%%%%%%%%%%%%%%%%%%%%%%%%%%%%%%%%%%%%%


\begin{thebibliography}{99}

%%%%%%%%%%%%%%%%%%%%%%%%%%%%%%%%%%%%%%%%%%%%%%%%%%%%%%%%%%%%%%%%%%%

\bibitem{GHZ89}
D.M.~Greenberger, M.A.~Horne, and A.~Zeilinger,
%``Going beyond Bell's theorem'',
in {\em Bell's Theorem, Quantum Theory, and Conceptions of the
Universe}, edited by M. Kafatos (Kluwer Academic, Dordrecht,
Holland, 1989), p. 69.

\bibitem{Mermin90a}
N.D. Mermin,
%``Simple unified form for the major no-hidden-variables theorems'',
%{\em Phys. Rev. Lett.} {\bf 65}, 27, 3373-3376 (1990).
Phys. Rev. Lett. {\bf 65}, 3373 (1990).

\bibitem{Bell64}
J.S. Bell,
%``On the Einstein-Podolsky-Rosen paradox'',
Physics (Long Island City, NY) {\bf 1}, 195 (1964).

\bibitem{Zukowski91}
M. \.{Z}ukowski,
%`Definite values for observables versus quantum predictions:
%A ``GHZ-like'' test',
Phys. Lett.~A {\bf 157}, 198 (1991).

\bibitem{CRB91}
R.K. Clifton, M.L.G. Redhead, and J.N. Butterfield,
%``Generalization of the Greenberger-Horne-Zeilinger algebraic proof
%of nonlocality'',
Found. Phys. {\bf 21}, 149 (1991).

\bibitem{CB97}
R. Cleve and H. Buhrman,
%``Substituting quantum entanglement for communication'',
Phys. Rev.~A {\bf 56}, 1201 (1997).

\bibitem{ZZHW98}
%M. \.{Z}ukowski, A. Zeilinger, M. A. Horne, and H. Weinfurter,
M. \.{Z}ukowski {\em et al.},
%``Quest for GHZ states'',
Acta Phys. Pol. A {\bf 93}, 187 (1998).

\bibitem{DVC00}
W. D\"{u}r, G. Vidal, and J. I. Cirac,
%``Three qubits can be entangled in two inequivalent ways'',
Phys. Rev.~A {\bf 62}, 062314 (2000).

\bibitem{EPR35}
A. Einstein, B. Podolsky, and N. Rosen,
%``Can quantum-mechanical description of physical reality
%be considered complete?'',
Phys. Rev. {\bf 47}, 777 (1935).

\bibitem{Mermin90b}
N.D. Mermin,
%``Extreme quantum entanglement in a
%superposition of macroscopically distinct states'',
Phys. Rev. Lett. {\bf 65}, 1838 (1990).

%%%%%%%%%%%%%%%%%%%%%%%%%%%%%%%%%%%%%%%%%%%%%%%%%%%%%%%%%%%%%%%%%%%
% 2-observer AVN with 2-qubit measurements
%%%%%%%%%%%%%%%%%%%%%%%%%%%%%%%%%%%%%%%%%%%%%%%%%%%%%%%%%%%%%%%%%%%

\bibitem{Cabello01a}
A. Cabello,
%``Bell's theorem without inequalities and without
%probabilities for two observers'',
Phys. Rev. Lett. {\bf 86}, 1911 (2001).

\bibitem{Cabello01b}
A. Cabello,
%```All versus nothing'' inseparability for two observers',
Phys. Rev. Lett. {\bf 87}, 010403 (2001).

\bibitem{CPZBZ03}
%Z.-B.~Chen, J.-W.~Pan, Y.-D.~Zhang, \v{C}.~Brukner, and A.~Zeilinger,
Z.-B.~Chen {\em et al.},
%``All-versus-nothing violation of local realism for two entangled photons'',
Phys. Rev. Lett. {\bf 90}, 160408 (2003).

%%%%%%%%%%%%%%%%%%%%%%%%%%%%%%%%%%%%%%%%%%%%%%%%%%%%%%%%%%%%%%%%%%%
% Experimental 2-observer AVN with 2-qubit measurements
%%%%%%%%%%%%%%%%%%%%%%%%%%%%%%%%%%%%%%%%%%%%%%%%%%%%%%%%%%%%%%%%%%%

\bibitem{CBPMD05}
%C. Cinelli, M. Barbieri, R. Perris, P. Mataloni, and F. De Martini,
C. Cinelli {\em et al.},
%``All-versus-nothing nonlocality test of quantum mechanics
%by two-photon hyperentanglement'',
Phys. Rev. Lett. {\bf 95}, 240405 (2005).

\bibitem{YZZYZZCP05}
%T. Yang, Q. Zhang, J. Zhang, J. Yin, Z. Zhao, M. \.{Z}ukowski, Z.-B.
%Chen, and J.-W. Pan,
T. Yang {\em et al.},
%``All-versus-nothing violation of local realism by two photon,
%four dimensional entanglement'',
Phys. Rev. Lett. {\bf 95}, 240406 (2005).

%%%%%%%%%%%%%%%%%%%%%%%%%%%%%%%%%%%%%%%%%%%%%%%%%%%%%%%%%%%%%%%%%%%

\bibitem{Lvovsky02}
A.I. Lvovsky,
%``Cabello's nonlocality and linear optics'',
Phys. Rev. Lett. {\bf 88}, 098901 (2002).

%%%%%%%%%%%%%%%%%%%%%%%%%%%%%%%%%%%%%%%%%%%%%%%%%%%%%%%%%%%%%%%%%%%
% 2-observer AVN with single-qubit measurements
%%%%%%%%%%%%%%%%%%%%%%%%%%%%%%%%%%%%%%%%%%%%%%%%%%%%%%%%%%%%%%%%%%%

\bibitem{Cabello05a}
A. Cabello,
%``Stronger two-observer all-versus-nothing violation of local realism'',
Phys. Rev. Lett. {\bf 95}, 210401 (2005).

\bibitem{Cabello05b}
A. Cabello,
%``Loophole-free Bell's experiment and two-photon all-versus-nothing
%violation of local realism'',
Phys. Rev. A {\bf 72}, 050101(R) (2005).

%%%%%%%%%%%%%%%%%%%%%%%%%%%%%%%%%%%%%%%%%%%%%%%%%%%%%%%%%%%%%%%%%%%
% Experimental 2-observer AVN with single-qubit measurements
%%%%%%%%%%%%%%%%%%%%%%%%%%%%%%%%%%%%%%%%%%%%%%%%%%%%%%%%%%%%%%%%%%%

\bibitem{VPMDB07}
%G. Vallone, E. Pomarico, P. Mataloni, F. De Martini, and V. Berardi,
G. Vallone {\em et al.},
%``Realization and characterization of a two-photon four-qubit linear cluster
%state'',
Phys. Rev. Lett. {\bf 98}, 180502 (2007).

%%%%%%%%%%%%%%%%%%%%%%%%%%%%%%%%%%%%%%%%%%%%%%%%%%%%%%%%%%%%%%%%%%%
% Hyper-entangled states encoding six qubits in two photon
%%%%%%%%%%%%%%%%%%%%%%%%%%%%%%%%%%%%%%%%%%%%%%%%%%%%%%%%%%%%%%%%%%%

\bibitem{BLPK05}
%J.~T.~Barreiro, N.~K.~Langford, N.~A.~Peters, and P.~G.~Kwiat,
J.T.~Barreiro {\em et al.},
%``Hyper-entangled photons'',
Phys. Rev. Lett. {\bf 95}, 260501 (2005).

%%%%%%%%%%%%%%%%%%%%%%%%%%%%%%%%%%%%%%%%%%%%%%%%%%%%%%%%%%%%%%%%%%%
% 6-photon states
%%%%%%%%%%%%%%%%%%%%%%%%%%%%%%%%%%%%%%%%%%%%%%%%%%%%%%%%%%%%%%%%%%%

\bibitem{ZGWCZYMSP06}
%Q. Zhang, A. Goebel, C. Wagenknecht, Y.-A. Chen, B. Zhao, T. Yang,
%A. Mair, J. Schmiedmayer, and J.-W. Pan,
Q. Zhang {\em et al.},
%``Experimental quantum teleportation of a two-qubit composite system'',
Nature Physics {\bf 2}, 678 (2006).

\bibitem{LZGGZYGYP07}
%C.-Y. Lu, X.-Q. Zhou, O. G\"{u}hne, W.-B. Gao, J. Zhang, Z.-S. Yuan,
%A. Goebel, T. Yang, and J.-W. Pan,
C.-Y. Lu {\em et al.},
%``Experimental entanglement of six photons in graph states'',
Nature Physics {\bf 3}, 91 (2007).

%%%%%%%%%%%%%%%%%%%%%%%%%%%%%%%%%%%%%%%%%%%%%%%%%%%%%%%%%%%%%%%%%%%

\bibitem{Cabello06a}
A. Cabello,
%``Two-player quantum pseudo-telepathy based on recent
%all-versus-nothing violations of local realism",
Phys. Rev. A {\bf 73}, 022302 (2006).

\bibitem{Cabello06b}
A. Cabello,
%``Bipartite Bell inequalities for hyperentangled states'',
Phys. Rev. Lett. {\bf 97}, 140406 (2006).

\bibitem{BDMVC06}
%M. Barbieri, F. De Martini, P. Mataloni, G. Vallone, and A. Cabello,
M. Barbieri {\em et al.},
%``Enhancing the violation of the Einstein-Podolsky-Rosen local realism
%by quantum hyperentanglement'',
Phys. Rev. Lett. {\bf 97}, 140407 (2006).

\bibitem{Gottesman96}
D. Gottesman,
%``Class of quantum error-correcting codes saturating the quantum Hamming bound'',
Phys. Rev.~A {\bf 54}, 1862 (1996).

%%%%%%%%%%%%%%%%%%%%%%%%%%%%%%%%%%%%%%%%%%%%%%%%%%%%%%%%%%%%%%%%%%%

\bibitem{VDM04}
M. Van den Nest, J. Dehaene, and B. De Moor,
%``Graphical description of the action of local Clifford transformations on graph states'',
Phys. Rev.~A {\bf 69}, 022316 (2004).

\bibitem{HEB04}
M. Hein, J. Eisert, and H.J. Briegel,
%``Multi-party entanglement in graph states,
Phys. Rev.~A {\bf 69}, 062311 (2004).

%%%%%%%%%%%%%%%%%%%%%%%%%%%%%%%%%%%%%%%%%%%%%%%%%%%%%%%%%%%%%%%%%%%

\bibitem{RB01}
R. Raussendorf and H.J. Briegel,
%``A one-way quantum computer'',
Phys. Rev. Lett. {\bf 86}, 5188 (2001).

\bibitem{HDERVB06}
%M. Hein, W. D\"{u}r, J. Eisert, R. Raussendorf, M. Van den Nest, and
%H. J. Briegel,
M. Hein {\em et al.},
%``Entanglement in graph states and its applications'',
in {\em Quantum Computers, Algorithms and Chaos},
edited by G. Casati {\em et al.}
%G. Casati, D.L. Shepelyansky, P. Zoller, and G. Benenti
(IOS Press, Amsterdam, 2006).

\bibitem{DP97}
D.P. DiVincenzo and A. Peres,
%``Quantum codewords contradict local realism'',
Phys. Rev.~A {\bf 55}, 4089 (1997).

\bibitem{SASA05}
%V. Scarani, A. Ac\'{\i}n, E. Schenck, and M. Aspelmeyer,
V. Scarani {\em et al.},
%``Nonlocality of cluster states of qubits'',
Phys. Rev.~A {\bf 71}, 042325 (2005).

%%%%%%%%%%%%%%%%%%%%%%%%%%%%%%%%%%%%%%%%%%%%%%%%%%%%%%%%%%%%%%%%%%%

\end{thebibliography}
\end{document}